\documentclass[aps,prl,preprint,superscriptaddress,showpacs]{revtex4-1}

\usepackage{graphicx}
\usepackage{dcolumn}
\usepackage{bm}

\begin{document}


\title{Observation of Strong Terahertz Radiation from a Liquid Water Line}

\author{L.-L. Zhang}
\affiliation{Beijing Advanced Innovation Center for Imaging
Technology and Key Laboratory of Terahertz Optoelectronics (MoE),
Department of Physics, Capital Normal University, Beijing 100048,
China}
\author{W.-M. Wang}\email{e-mail: weiminwang1@126.com}
\affiliation{Beijing National Laboratory for Condensed Matter
Physics, Institute of Physics, CAS, Beijing 100190,
China}\affiliation{Beijing Advanced Innovation Center for Imaging
Technology and Key Laboratory of Terahertz Optoelectronics (MoE),
Department of Physics, Capital Normal University, Beijing 100048,
China}\affiliation{Songshan Lake Materials Laboratory, Dongguan,
Guangdong 523808, China}

\author{T. Wu}
\affiliation{Beijing Key Laboratory for
Precision Optoelectronic Measurement Instrument and Technology,
School of Optoelectronics, Beijing Institute of Technology, Beijing
100081, China}

\author{S.-J. Feng}
\affiliation{Beijing Key Laboratory for Precision Optoelectronic
Measurement Instrument and Technology, School of Optoelectronics,
Beijing Institute of Technology, Beijing 100081, China}

\author{K. Kang}
\affiliation{Beijing Advanced Innovation Center for Imaging
Technology and Key Laboratory of Terahertz Optoelectronics (MoE),
Department of Physics, Capital Normal University, Beijing 100048,
China}

\author{C.-L. Zhang}
\affiliation{Beijing Advanced Innovation Center for Imaging
Technology and Key Laboratory of Terahertz Optoelectronics (MoE),
Department of Physics, Capital Normal University, Beijing 100048,
China}

\author{Y. Zhang}
\affiliation{Beijing Advanced Innovation Center for Imaging
Technology and Key Laboratory of Terahertz Optoelectronics (MoE),
Department of Physics, Capital Normal University, Beijing 100048,
China}

\author{Y.-T. Li}
\affiliation{Beijing National Laboratory for Condensed Matter
Physics, Institute of Physics, CAS, Beijing 100190,
China}\affiliation{Songshan Lake Materials Laboratory, Dongguan,
Guangdong 523808, China}

\author{Z.-M. Sheng}
\affiliation{SUPA, Department of Physics, University of Strathclyde,
Glasgow G4 0NG, United Kingdom} \affiliation{Key Laboratory for
Laser Plasmas (MoE) and School of Physics and Astronomy, Shanghai
Jiao Tong University, Shanghai 200240, China}\affiliation{IFSA
Collaborative Innovation Center, Shanghai Jiao Tong University,
Shanghai 200240, China}\affiliation{Tsung-Dao Lee Institute,
Shanghai Jiao Tong University, Shanghai 200240, China}

\author{X.-C. Zhang}
\affiliation{The Institute of Optics, University of Rochester,
Rochester, New York 14627, USA} \affiliation{Beijing Advanced
Innovation Center for Imaging Technology and Key Laboratory of
Terahertz Optoelectronics (MoE), Department of Physics, Capital
Normal University, Beijing 100048, China}

\date{\today}

\begin{abstract}
Terahertz radiation generation from liquid water has long been
considered to be impossible due to strong absorption. A few very
recent works reported terahertz generation from water, but the
mechanism is not clear and the efficiency demands to be enhanced. We
show experimentally that strong single-cycle terahertz radiation
with field strength of $\rm 0.2~MV cm^{-1}$ is generated from a
water line/column of $\sim 200 \mu m$ in diameter irradiated by a mJ
femtosecond laser beam. This strength is 100-fold higher than that
produced from air. We attribute the mechanism to the
laser-ponderomotive-force-induced current with the symmetry broken
around the water-column interface. This mechanism can explain our
following observations: the radiation can be generated only when the
laser propagation axis deviates from the column center; the
deviation determines its field strength and polarity; it is always
p-polarized no matter whether the laser is p- or s-polarized. This
study provides a simple and efficient scheme of table-top terahertz
sources based on liquid water.
\end{abstract}


\maketitle

Achieving table-top terahertz (THz) sources with high field strength
and broad bandwidth is an outstanding issue in THz science. Such
sources can find applications in material research
\cite{Ferguson,Clough}, biomedical imaging \cite{Mittleman},
non-destructive detection \cite{Tonouchi}, and THz-field matter
interactions \cite{THz-phy3,THz-CP}. Previous studies have
demonstrated THz generation from solids \cite{Jin,Fulop,Gopal,Liao}
and gases
\cite{Cook,Kress,Xie,NatPh,WuHC,Kim,THz_OE,Babushkin,WL_Scaling,Andreeva,THz_PRL2017}
via different mechanisms. However, THz generation from liquid, in
particular water, has long been considered impossible because of its
strong absorption of THz radiation. On the other hand, water exists
in most biological systems and hence THz radiation generated from
liquid water may carry some information of these systems. Therefore,
how to generate THz radiation from water is fundamental challenge
for both basic and applied research. In 2017, two groups reported
THz emission from liquid water \cite{NC2017_water,APL2017_water}.
When an intense laser beam of tens of mJ was focused on liquid water
in a cuvette, extreme broadband THz radiation was generated
\cite{NC2017_water}, where it is considered that laser spectral
broadening played a key role. In the other work
\cite{APL2017_water}, when a mJ laser beam irradiated a water film
with the thickness $\sim 200~\rm\mu m$, THz radiation was produced
with 1.8 times higher strength than that produced from air. So far,
the THz radiation mechanism in water has not yet been well clarified
and the yield efficiency demands to be further enhanced.

Here, we demonstrate experimentally that the efficiency can be
enhanced by three orders of magnitude when a water column with the
diameter $\sim 200~\rm\mu m$ is adopted. With a mJ femotsecond laser
beam, the THz field strength can reach $\rm 0.2~MV cm^{-1}$ which is
as high as generated via the standard two-color laser scheme in air
\cite{Cook,Kress}. To explain our result, we propose that the THz
radiation originates from a net current formed due to the presence
of the column interface. The laser self-focusing in water causes a
plasma to be produced. The laser ponderomotive force forms positive
and negative currents distributed on two sides of the laser
propagation axis, respectively. The symmetry of the two currents can
be broken provided the laser axis deviates from the water column
center. As the deviation grows, the net current and resulting THz
radiation will be strengthened. This mechanism implies that the THz
polarization is on the column cross-section plane and its strength
scales linearly with the laser energy. These are verified by our
experiments and particle-in-cell (PIC) simulations.

\begin{figure}[htbp]
\includegraphics[width=6in]{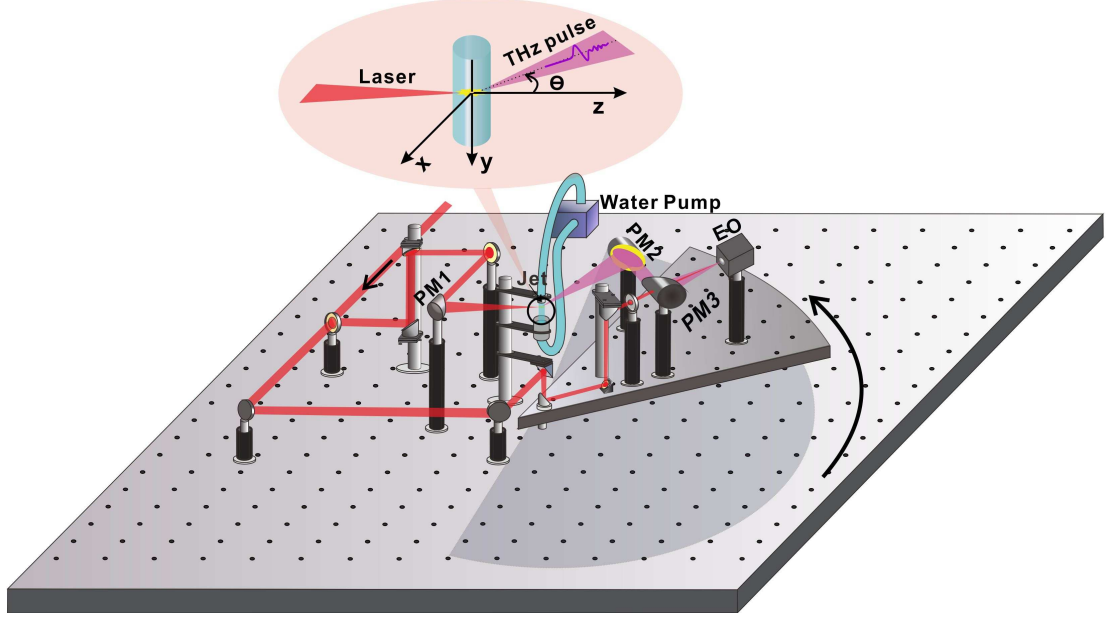}
\caption{\label{fig:epsart}Experimental setup, where PM1-3 are
parabolic mirrors and EO is electro-optical detection. The inset
illustrates the geometry of the laser interaction with the water
column, where THz pulses can be detected at an angle $\theta$
rotating from the laser axis ($z$) in the incident plane $xoz$.}
\end{figure}

\emph{Experimental setup.}$-$ Figure 1 shows a schematic of our
experiment, where a laser beam is incident along the $+z$ direction
and the water column axis is along the $y$ direction. The laser beam
is delivered from a Ti:Sapphire amplifier (Spitfire, Spectra
Physics) with a central wavelength 800nm, pulse duration 100fs, and
repetition rate 1kHz. It is split into pump and probe beams with
controllable time delay. The pump beam is focused by an off-axis
parabolic mirror (PM1) with 1-inch equivalent focal length. The
polarization of the pump beam is linear and its orientation can be
rotated through a half-wave plate. A liquid geyser with a pressure
of 0.1MPa creates a free-flowing water column with the diameter
$\sim 200~\rm\mu m$ near the tip of the geyser. The water column is
located around the focusing plane of the pump beam and can precisely
move along the $x$ direction (equivalent to the shift of the laser
propagation axis). Here, we fix the coordinate on the column and set
the column center as the origin, as shown in the inset.

The THz pulse is collimated and refocused by two parabolic mirrors.
Filters are placed in the THz path to block the residual laser beam.
The probe laser beam passes through a pair of climbing mirrors, is
focused by a 125mm convex lens, and co-propagates with the THz pulse
which has passed through a hole drilled on the back of the parabolic
mirror (PM3). The collection portion is installed on a platform
which can be rotated around the water column to detect the THz pulse
at an angle $\theta$ (positive: anti-clockwise) with respect to the
laser incident direction. To reduce the user intervention and the
experimental error, we minimize the optical path difference between
the pump and probe beam arms of rotation. The THz fields resolved
traces are obtained through electro-optic (EO) sampling with a 3 mm
thick $\langle110\rangle$-cut ZnTe crystal as the detector
\cite{APL_Zhang1995}. In our experiments, the laser beam is taken as
2 mJ energy, p-polarization (along the $x$ direction), the laser
propagation axis is displaced $60~\rm\mu m$ ($x_L=60 ~\rm\mu m$)
from the water column center, and the THz pulse is collected at
$\theta =0^\circ$, except in Fig. 4.

\begin{figure}[htbp]
\includegraphics[width=6in]{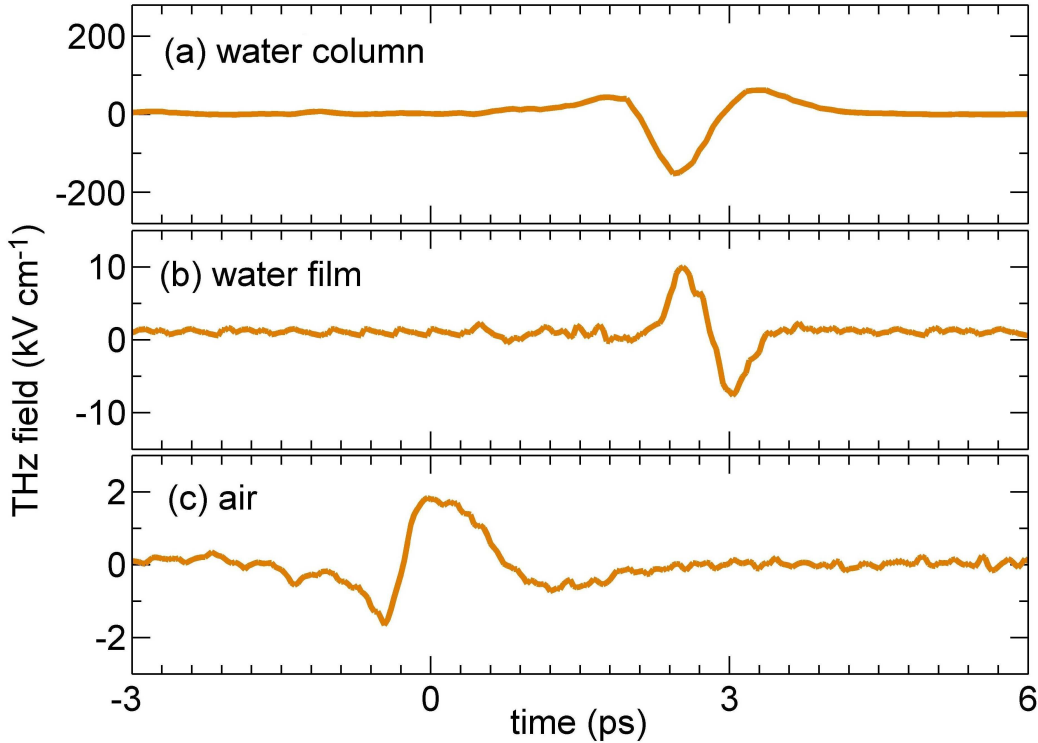}
\caption{\label{fig:epsart} THz pulses generated from (a) water
column with the normal laser incidence and $x_L=60 ~\rm\mu m$, (b)
water film with the laser incident angle of $60^\circ$, and (c) air
with the normal laser incidence. The THz pulses are detected by
electro-optic sampling and collected at $\theta=0^\circ$.}
\end{figure}

\emph{Demonstration of THz generation.}$-$ Figure 2(a) shows the
waveform of the THz pulse generated from the water column. As
comparison, the ones from water film and air irradiated by the same
laser beam are also displayed in Figs. 2(b) and 2(c). The THz pulse
from the water column has a field strength about $\rm 0.2~MV
cm^{-1}$, 20-fold and 100-fold higher than the one from the water
film and air, respectively. The THz strength is as high as the one
with the standard two-color laser scheme in air \cite{Cook,Kress}
even though a one-color laser beam is used here. Note that the
strength can be further enhanced when the THz pulse is collected at
$\theta$ of $40^\circ-60^\circ$ rather than $\theta=0^\circ$ [this
will be shown in Fig. 4(c)]. In Fig. 2(b) we take a 200-$\rm \mu
m$-thick and 5-mm-wide water film, which is produced by a jet nozzle
with polished sapphire surfaces (Sirah, Germany). The laser incident
angle is taken as $60^\circ$ to optimize the THz strength, in
particular, nearly no THz generation with the laser normal incidence
\cite{APL2017_water}. However, in the water column case, the normal
laser incidence along the $+z$ direction is always taken in our
experiments and efficient THz generation is observed (note that the
laser beam with a self-focusing intensity $\sim 10^{15}~\rm W
cm^{-2}$ ionizes the water column to be plasma and then the beam can
propagate along its incident direction since the plasma refractive
index approaches 1 in our case). This suggests that there are
different generation mechanisms in the two cases [different THz
strength scaling are also observed as shown in Fig. 4(a)]. Here, we
focus on the water column case and exploration of the mechanism in
the water film is beyond the scope of this work.

\emph{Mechanism.}$-$ The experimental and PIC-simulation results
shown in Fig. 3 suggest that the mechanism can be explained as the
laser-ponderomotive-force-induced current with the symmetry broken
around the column interface. Figure 3(a) shows that the THz pulses
have nearly the same amplitude and opposite field signs when the
laser axis deviates from the column center by $+60\mu m$ and $-60\mu
m$ ($x_L=\pm 60\mu m$), respectively. While the laser axis is at the
column center ($x_L=0$), virtually no THz pulse is generated, as
seen in Fig. 3(b). This figure also shows that the THz strengths
have the same absolute value and opposite signs at the two points
$\pm x_L$. As $|x_L|$ is increased, the amplitude is first enhanced
and then lowered. The amplitude peaks appear around $x_L=\pm (60\mu
m\sim70\mu m)$. These are in agreement with our PIC simulation
results shown by the line in Fig. 3(b), which are explained below.

Our simulations are performed with the KLAPS code \cite{KLAPS}, in
which we adopt the same parameters of the water column and laser
(energy, duration, and polarization) as in the experiments.
Considering that the laser self-focusing in water should be stronger
than in air, we assume that the laser beam in the water column has
the spot radius $w_0=30 ~\rm \mu m$. Then, the corresponding
intensity is $1.5\times10^{14}~\rm W cm^{-2}-1.7\times10^{15}~\rm W
cm^{-2}$ when the laser energy varies within $0.2~\rm mJ-2.4~\rm
mJ$. In our simulations, the laser energy is taken as 2mJ
($1.2\times10^{15}~\rm W cm^{-2}$) except Fig. 4(a). With such laser
intensity, plasma is quickly produced by the laser beam via field
ionization. No net current can be formed via the ionization since
the symmetry of the ionization by a one-color 800nm laser beam is
not broken \cite{Kim,THz_OL}.

\begin{figure}[htbp]
\includegraphics[width=6in]{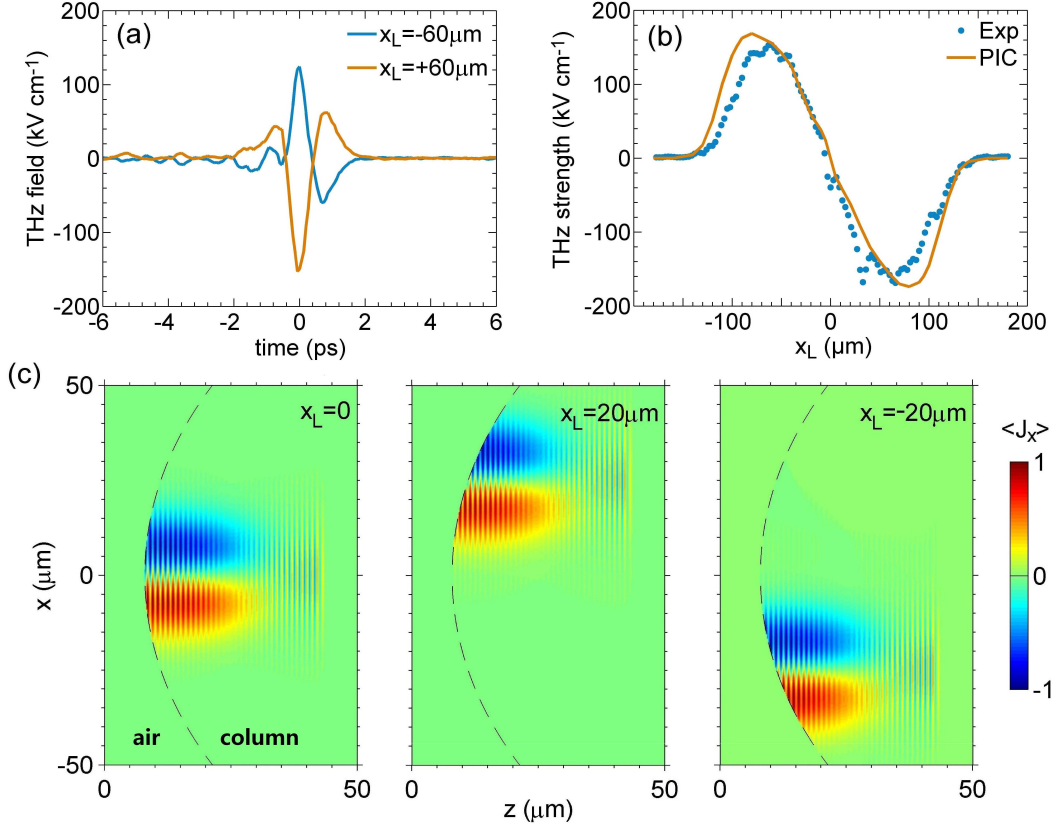}
\caption{\label{fig:epsart}(a) THz pulses observed in the
experimenters with $x_L=\pm 60 \mu m$. (b) THz strength as a
function of $x_L$, where experimental and PIC results are shown by
dots and line, respectively. (c) PIC results of quasi-static
currents $\langle J_x \rangle$ with different $x_L$, where the
currents are obtained by temporally averaging $J_x$ over one laser
cycle and the broken line in each plots marks the column interface.}
\end{figure}

On the other hand, our simulations show in Fig. 3(c) that net
currents can be formed in the laser interactions with the
water-column plasma. We examine the quasi-static currents $\langle
J_x\rangle$, $\langle J_y\rangle$, and $\langle J_z\rangle$,
respectively, where these currents are obtained by temporally
averaging $J_x$, $J_y$, and $J_z$ over one laser cycle. Here, the
laser polarization is along the $x$ direction (p-polarization). One
can see in Fig. 3(c) that the total/net currents $\sum\langle
J_x\rangle\neq0$ unless $x_L=0$, where $\sum$ means spatial
summation. When $x_L=0$, the positive and negative currents are
symmetrically distributed, therefore, the net current is zero. When
$x_L=20~\rm\mu m$, the positive current is distributed within a
larger area than the negative one, therefore, the net current is
positive. While $x_L=-20~\rm\mu m$, the positive current is
distributed within a smaller area, therefore, the net current is
negative. In addition, the net currents $\sum\langle J_y\rangle$ and
$\sum\langle J_z\rangle$ remain zero with any $x_L$, as in usual
cases without special target interfaces.

The net current is formed in the $x$ direction because of the column
interface in the laser incident plane as well as the spatial
non-uniform of laser intensity. With a Gaussian laser beam, the
laser ponderomotive force \cite{P.Gibbon} pushes the plasma
electrons away from the laser axis. Hence, the quasi-static current
is negative on the upper($x>x_L$) and positive on the lower
($x<x_L$). Around the column interface, the electrons pushed by the
ponderomotive force are pulled back by the plasma ions due to strong
charge-separation fields. This prevents the electrons from escaping
away from the interface, constraining the current near the
interface. Hence, the interface breaks the symmetry between the
positive and negative currents. The area difference between the
positive and negative currents can be estimated with
\begin{eqnarray}
\Delta S = 2f(x_L)-f(x_L+w)-f(x_L-w),
\end{eqnarray}
where $f(x)=[x\sqrt{R^2-x^2}+R^2\arcsin(x/R)]/2$, $R$ is the column
radius and $w$ is the efficient width of the laser beam ($w$ at the
order of $w_0$). Obviously, $|\Delta S|$ grows with increasing
$|x_L|$. Provided $x_L$ is replaced by $-x_L$, the absolute value of
$\Delta S$ remains constant, but its sign is reversed. This can
explain the experimental and PIC results within $|x_L|<x_L^{opt}$
shown in Fig. 3(b) (THz peaks at $\pm x_L^{opt}$), since
$J_{net}\propto \Delta S$ and $E_{THz} \propto J_{net}$
\cite{THz_OE,THz_PRL2017}. Note that in Figs. 3(b),4(a),4(b) we
present the net-current strengths obtained from our PIC simulations.
Similar to the two-color scheme in air \cite{Kim,Babushkin}, THz
pulses can be generated once net currents are formed in plasma due
to the plasma modulation \cite{THz_OE}, which causes single-cycle
waveforms of the THz pulses [see Figs. 2(a),3(a)]. Note that we have
discussed $\Delta S$ in the laser-entrance side, which can directly
be applied in the laser-exit side since the laser beam is normally
incident and the water plasma density is low.

Figure 3(b) also shows that there are optimized values of $x_L$ for
the THz field strength. This is because the zone of laser-water and
laser-plasma interactions becomes too small if $|x_L|$ is taken as a
large value, which limits the THz generation. According to our PIC
simulations, $x_L^{opt}$ depends on the laser spot radius $w_0$:
$x_L^{opt}\simeq 92\mu m$ with $w_0=15\mu m$, $x_L^{opt}\simeq 80\mu
m$ with $w_0=30\mu m$ [see Fig. 3(b)], and $x_L^{opt}\simeq 70\mu m$
with $w_0=45\mu m$. Based on the simulation results, we could
roughly summarize as $x_L^{opt} \simeq R-2w_0/3$. This value varies
slightly with the laser energy within the range of 0.2mJ to 2.4mJ.

\emph{THz strength scaling and polarization.}$-$The THz field
strength scaling with the laser intensity or energy is determined by
the ponderomotive force. In a laser field, motion of an electron is
governed by the Hamiltonian $ H=m c^2\gamma-e\varphi$, where ${\bf
p}$ and $\gamma=\sqrt{1+({\bf p}/mc)^2}$ are the momentum and
relativistic factor, respectively, $e$ and $m$ are the electron
charge and mass, respectively, $c$ is the light speed in vacuum, and
$\varphi$ is the scalar potential generated due to the plasma
response. Taking the spatial derivative of $H$, one can obtain
$dp_z/dt=\partial (e\varphi-m c^2\gamma) /
\partial z$ and $d({\bf p}_\bot-e{\bf A}/c)/dt = \nabla_\bot
(e\varphi-m c^2\gamma)$, where ${\bf A}$ is the laser vector
potential. We consider a plasma with the plasma oscillating
frequency $\omega_p=\sqrt{4\pi e^2 n_e/m}$ much lower than the laser
frequency $\omega$, where $n_e$ is the plasma density. Note that the
THz pulse frequency, which is roughly at $\omega_p/2\pi$
\cite{THz_OE,THz_PRL2017}, is close to 1 THz according to Fig. 2(a).
Therefore, it can be assumed that any physical quantity $Q$ in this
laser-plasma system can be divided into a fast varying part and a
slowly varying part, i.e., $Q=Q^f+\langle Q \rangle$, where $Q^f$
varies at the order of $\omega$, $\langle Q \rangle$ at the order of
$\omega_p$, $\langle Q \rangle=\int_0^T Q dt /T$, and
$T=2\pi/\omega$ is the laser cycle. The fast varying part of the
momentum satisfies $dp_z^f/dt=-m c^2
\partial \gamma^f/\partial z$ and $d({\bf p}_\bot^f-e{\bf
A}/c)/dt=0$. The slowly varying part satisfies $d \langle {\bf p
}_\bot \rangle/dt=e\nabla_\bot\varphi-m c^2\nabla_\bot\langle \gamma
\rangle$, where the first term on the right hand is the
electrostatic force and the second is the ponderomotive force ${\bf
F}_{p}$. In our case with $\omega_p\ll\omega$, basically $|\langle
{\bf p} \rangle| \ll |{\bf p}^f|$ and $|e\nabla_\bot\varphi| \ll
F_{p}$. Therefore, $\gamma\simeq 1+e^2{\bf A}^2/2m^2c^4$
\cite{Wang_LPFA} and $d \langle {\bf p }_\bot \rangle/dt \simeq {\bf
F}_{p}= -e^2 \nabla_\bot \langle{\bf A}^2 \rangle /2mc^2$. By
applying $\langle {\bf J}_\bot \rangle=-en_e \langle {\bf p }_\bot
\rangle/m$ in a non-relativistic case, the quasi-static current
induced by the ponderomotive force is given by
\begin{eqnarray}
\langle \frac{\partial {\bf J}_\bot}{\partial t} \rangle \simeq
\frac{e^3n_e}{2m^2 c^2}\nabla_\bot \langle{\bf A}^2 \rangle.
\end{eqnarray}
This equation gives $\langle \partial {\bf J}_\bot/\partial t
\rangle \propto A_{0}^2/w_0^2 \propto \varepsilon_{laser}/w_0^2$,
where we consider a Gaussian beam with $\nabla_\bot \langle{\bf A}^2
\rangle \sim A_{0}^2/w_0^2$ and the lase energy $\varepsilon_{laser}
\propto A_{0}^2$. According to ${\bf E}_{THz} \propto \langle
\partial {\bf J}_\bot/\partial t \rangle$
\cite{THz_OE,THz_PRL2017}, one can obtain:
\begin{eqnarray}
E_{THz} \propto \frac{\varepsilon_{laser}}{w_0^2}.
\end{eqnarray}

This linear scaling of the THz field strength with the laser energy
is verified by our experimental and PIC results as shown in Fig.
4(a). Note that this scaling is different from that in the water
film case \cite{APL2017_water}, in which $E_{THz} \propto
\sqrt{\varepsilon_{laser} }$. Equation (3) also suggests that the
THz strength is decreased with the laser spot radius $w_0$ in the
water plasma. This is difficult to examine by experiments since
$w_0$ is mainly determined by the laser self-focusing in water. Our
PIC simulation results roughly follow the scaling with $1/w_0^2$.
For example, the net currents with $w_0=15\mu m$ is 3-6 times
(varying with $x_L$) of those with $w_0=30\mu m$ when the laser
intensity is fixed. The deviation from the predicted value 4 could
be explained as $\Delta S$ also depends on $w_0$ and $x_L$. Note
that the plasma density $n_e$ is nearly unchanged when the laser
energy is taken between 0.2mJ and 2.4mJ with the corresponding
intensity $2\times 10^{14}~\rm W/cm-1.7\times 10^{15}~\rm W
cm^{-2}$. In this intensity range, the first order of complete
ionization occurs for both oxygen and hydrogen, but the second order
of ionization of the oxygen can be ignored because it requires an
intensity above $2\times 10^{15}~\rm W cm^{-2}$.

The ponderomotive-induced current given in equation (2) is symmetric
in any transverse direction, e.g., it is negative at $y>0$ and
positive at $y<0$, which exactly counteract each other. Hence, no
net current can be formed in a transverse direction, except in the
$x$ direction. In this direction, the symmetry of the current can be
broken by the water-column interface, as shown in Fig. 3(c). As a
result, the THz polarization is always along the $x$ direction
(p-polarized), no matter whether the laser beam is taken as
p-polarization or not. This is verified by our experiments, as shown
in Fig. 4(b). We record the transverse components of the THz
electric field by electro-optic sampling and then obtain the
polarization trajectory by recomposing the THz fields. When we
change the laser polarization angle from $0^\circ$ (p-polarized) to
$90^\circ$ (s-polarized), the THz pulse keeps p-polarized (more
results with different polarization angles are shown in Supplemental
Material). These experimental results are reproduced by our PIC
simulations.

\begin{figure}[htbp]
\includegraphics[width=6in]{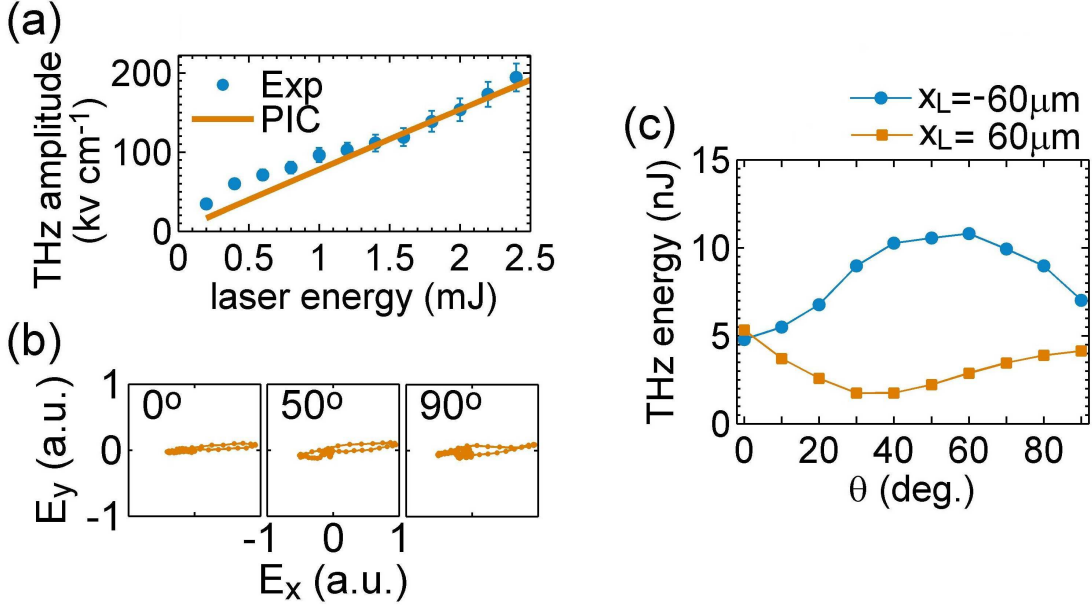}
\caption{\label{fig:epsart} (a) THz amplitude as a function of the
laser energy, where experimental results are shown by dots and PIC
results by the line. (b) Polarization trajectories of the $x$ and
$y$ components of THz fields obtained experimentally, where three
laser polarization angles $0^\circ$, $50^\circ$, and $90^\circ$ are
taken, respectively. (c) The THz energy as a function of $\theta$
observed in our experiments with $x_L=\pm 60 \mu m$, where the
detector is located at $x<0$.}
\end{figure}

Figure 4(c) shows the angular distribution of the THz pulses in the
range of $0^\circ-90^\circ$. The THz pulses are stronger with
$x_L=-60 \rm \mu m$ than those with $x_L=60 \rm \mu m$. This is
because the detector is rotated with $\theta>0$ and located at $x<0$
(see Fig. 1), the THz pulses with $x_L=60 \rm \mu m$ need to pass
through a longer distance of both water and plasma, hence, they are
more strongly absorbed. With $x_L=-60 \rm \mu m$ the peak angles
appear around $40^\circ-60^\circ$. In this case, the pulses
propagate mainly in the plasma towards the detector. Considering
that the net current is along the $x$ direction, the strongest
emission from the current should be at $\theta=0^\circ$ and it
weakens with increasing $\theta$. On the other hand, with
$\theta=0^\circ$ the THz pulse propagates the longest distance in
the plasma and it is most strongly absorbed and scattered by the
plasma. The propagation distance and the absorption decreases with
increasing $\theta$. The two factors cause the strongest THz pulses
to be observed at $40^\circ-60^\circ$. These factors can also
explain the energy decline from $\theta=0^\circ$ to $30^\circ$ in
the case with $x_L=60 \rm \mu m$. However, paths of the pulses
detected at larger $\theta$ are difficult to obtained because they
are affected by scattering and refraction at plasma-water and
water-air boundaries and they significantly deviate from the initial
emission direction. Finally, according to Fig. 4(c) we calculate the
THz yield efficiency to be above $6\times10^{-5}$, which is as high
as that with the two-color scheme pumped by 800nm lasers
\cite{WL_Scaling,Wang_TJ}.

In summary, we have proposed an efficient scheme to generate
liquid-water-based THz radiation with a single laser beam, where the
field strength and yield efficiency are as high as the standard
two-color laser scheme in gases. Our experiments have shown that a
water column irradiated by a 800 nm one-color laser beam of 2 mJ can
emit broadband THz radiation with the strength of $\rm 0.2~MV
cm^{-1}$, two orders of magnitude higher than one from air or a
water film. A laser-ponderomotive-force-induced current model has
been proposed to explain the THz generation mechanism. The model
predicts the dependence of the THz generation on laser energy,
polarization, as well as the deviation between the laser axis and
the column center, which has been verified by our experiments and
PIC simulations. In particular, the THz field strength and even
polarity can be controlled by the deviation.

\begin{acknowledgments}
This work was supported by National Key R\&D Program of China (Grant
No. 2018YFA0404801), National Natural Science Foundation of China
(Grants No. 11775302 and 11721091), and Science Challenge Project of
China (Grant No. TZ2016005). Z.-M. S. acknowledges the support of a
Leverhulme Trust Research Grant at the University of Strathclyde.
X.-C. Z. was also partially sponsored by the Army Research Office
and was accomplished under Grant No. US ARMY W911NF-17-1-0428. We
thank Prof. David R. Jones for useful discussion.
\end{acknowledgments}

%
%
%
%
%
%
%
%
%
%

\end{document}